\documentclass[12pt,a4paper]{article}
\usepackage{graphicx}
\usepackage[T1]{fontenc}
\usepackage[utf8]{inputenc}
\usepackage{textcomp}
\usepackage[sc,osf]{mathpazo}
\usepackage{a4wide}  
\usepackage{latexsym,amsthm,amsfonts,amsmath,mathrsfs,amssymb}
\usepackage{dsfont}
\usepackage{accents}
\usepackage[nosort]{cite}
\usepackage{booktabs} 
\usepackage{rotating}
\usepackage[unicode,implicit]{hyperref}
\hypersetup{%
  pdftitle    = {Covariant generalized conserved charges of General Relativity}
  pdfkeywords = {conserved charges, gravity, general relativity}
  pdfauthor   = {Carmen G\'omez-Fayr\'en, Patrick Meessen and Tom\'as Ort\'{\i}n},
  plainpages  = true,
  colorlinks  = true,
  citecolor   = blue,
  urlcolor    = red,
  linkcolor   = black
}
\newcommand{\hepth}[1]{{\tt
\href{http://www.arXiv.org/abs/hep-th/#1}{hep-th/#1}}}
\newcommand{\grqc}[1]{{\tt
\href{http://www.arXiv.org/abs/gr-qc/#1}{gr-qc/#1}}}

\newcommand{\arxiv}[1]{{\tt arXiv:\href{http://www.arXiv.org/abs/#1}{#1}}}
\newcommand{\doi}[1]{{\tt DOI:\href{http://dx.doi.org/#1}{#1}}}

\allowdisplaybreaks

\makeatletter
\@addtoreset{equation}{section}
\makeatother

\begin{document}

\begin{flushright}
\small
IFT-UAM/CSIC-23-076\\
July 8\textsuperscript{th}, 2023\\
\normalsize
\end{flushright}

\vspace{1cm}

\begin{center}

  {\Large {\bf Covariant generalized conserved charges of\\[.5cm]
      General Relativity}}

\vspace{1.2cm}

\renewcommand{\thefootnote}{\alph{footnote}}

{\sl\large Carmen G\'omez-Fayr\'en,$^{1,}$}\footnote{Email: {\tt carmen.gomez-fayren[at]estudiante.uam.es}}
{\sl\large Patrick Meessen$^{2,3}$}\footnote{Email: {\tt meessenpatrick[at]uniovi.es}}
{\sl\large  and Tom\'{a}s Ort\'{\i}n}$^{1,}$\footnote{Email: {\tt Tomas.Ortin[at]csic.es}}

\setcounter{footnote}{0}
\renewcommand{\thefootnote}{\arabic{footnote}}

\vspace{.6cm}

${}^{1}${\it Instituto de F\'{\i}sica Te\'orica UAM/CSIC\\
C/ Nicol\'as Cabrera, 13--15,  C.U.~Cantoblanco, E-28049 Madrid, Spain}

\vspace{0.2cm}

${}^{2}${\it HEP Theory Group, Departamento de F\'{\i}sica, Universidad de Oviedo\\
  Calle Leopoldo Calvo Sotelo 18, E-33007 Oviedo, Spain}\\

\vspace{0.2cm}

${}^{3}${\it Instituto Universitario de Ciencias y Tecnolog\'{\i}as Espaciales
  de Asturias (ICTEA)\\ Calle de la Independencia, 13, E-33004 Oviedo, Spain}

\vspace{1cm}


{\bf Abstract}
\end{center}
\begin{quotation}
  {\small Motivated by the current research of generalized symmetries and the
    construction of conserved charges in pure Einstein gravity linearized over
    Minkowski spacetime in Cartesian coordinates, we investigate, from a
    purely classical point of view, the construction of these charges in a
    coordinate- and frame-independent language in order to generalize them
    further. We show that all the charges constructed in that context are
    associated to the conformal Killing-Yano 2-forms of Minkowski
    spacetime. Furthermore, we prove that those associated to closed conformal
    Killing-Yano 2-forms are identical to the charges constructed by Kastor
    and Traschen for their dual Killing-Yano $(d-2)$-forms. We discuss the
    number of independent and non-trivial gravitational charges that can be
    constructed in this way.  }
\end{quotation}

\newpage
\pagestyle{plain}

\tableofcontents


\section{Introduction}

The definition of the conserved charges of the solutions of gravitational
theories is a fascinating topic of research that touches the foundations of
our current understanding of the gravitational field.\footnote{See
  Refs.~\cite{Arnowitt:1962hi,Abbott:1981ff,Lee:1990nz,Wald:1993nt,Iyer:1994ys,Barnich:1994db,Barnich:2000zw,Barnich:2001jy,Kastor:2004jk}
  for a somewhat \textit{ad hoc} selection of milestones in the history of
  this field of research.} It is our current understanding, at least at a
classical level, that the conserved charges of a spacetime that asymptotes to
another one which is treated as the vacuum, (this is the definition of an
isolated system in the gravitational setting) are associated to the isometries
of that vacuum spacetime just as the constants of motion of a point particle
moving in that vacuum spacetime are. In a sense, very far away from the strong
gravity region where the system whose charges we want to compute lies, that
system can be viewed as such a particle moving in the vacuum spacetime and we
know how to define the charges of that particle.

Mathematically, this associates the conserved charges of gravitational systems
to the Killing vectors of the vacuum spacetime. However, it has long been
known that particles and fields evolving in a spacetime may have other
conserved charges as well. The simplest example is provided by massless
particles, which have conserved charges associated to the conformal Killing
vectors of the spacetime. There are charges associated to Killing-Yano forms
(or tensors) \cite{kn:Yano,Cariglia:2003kf} as well and, in general, to
conformal Killing-Yano $p$-forms \cite{Jezierski:1994xm}.\footnote{For a
  review with many references focused on the construction of conserved
  quantities of particles and fields evolving in spacetimes admitting
  conformal Killing-Yano $p$-forms see Ref.~\cite{Frolov:2017kze} and the more
  recent Ref.\cite{Lindstrom:2022nrm}. Further and later results in this area
  can be found in
  Refs.~\cite{Lindstrom:2021qrk,Lindstrom:2022qjx,OkanGunel:2023yzv,Lindstrom:2021dpm,Lindstrom:2022iec}.}

According to the preceding discussion, it should be possible to define
conserved charges to gravitating isolated systems associated to the conformal
Killing-Yano $p$-forms of the asymptotic vacuum spacetime and, indeed, in
Ref.~\cite{Kastor:2004jk} Kastor and Traschen found a definition of off-shell
conserved charges associated to the Killing-Yano $p$-forms of the asymptotic
vacuum spacetime, which did not include the more general conformal
Killing-Yano $p$-forms, though.

More recently and with different goals, the definition of the conserved
charges of quantum gravitational systems has been discussed in
Refs.~\cite{Benedetti:2021lxj,Benedetti:2022zbb,Hinterbichler:2022agn,Benedetti:2023ipt}. In
these references and, without any reference to classical symmetries of the
vacuum spacetime, conserved 2- and $(d-2)$-form charges were constructed in
terms of the Riemann tensor linearized over Minkowski spacetime (playing the
role of vacuum). Since this construction is not connected to previous works in
the classical setting and may include new charges not considered so far in it,
it is interesting to understand them better from that point of view.

Thus, in this paper, we want to review the definitions of conserved
gravitational charges made in
Refs.~\cite{Benedetti:2021lxj,Benedetti:2022zbb,Hinterbichler:2022agn,Benedetti:2023ipt}
and relate them to already existing definitions in the classical realm. We
will refer mostly to the construction in $d$ dimensions made in
Ref.~\cite{Benedetti:2023ipt}.  We would like to stress that our approach is
purely classical and we will not be concerned with the implications of our
results in the context of the algebraic approach to Quantum Gravity.

We will start by reviewing in Section~\ref{sec-definitions} the definitions
made in Ref.~\cite{Benedetti:2023ipt} to reformulate them in a coordinate- and
frame-independent way. As we are going to see, it is natural and unavoidable
to generalize those definitions to $p$-form charges and to consider also the
$p$-form charged defined by Kastor and Traschen. Fortunately, we will be able
to prove a Lemma relating many of those charges, simplifying their
classification and interpretation. In Section~\ref{sec-conservationofQ} we
prove the conservation of those charges under different assumptions concerning
the properties of the $p$-form parameters used. Essentially we will be able to
extend the definition of the Kastor-Traschen charges to the case in which the
parameters are more general conformal Killing-Yano $p$-forms. In
Section~\ref{sec-AD} we apply the results of the previous section to
asymptotically-flat spacetimes using the conformal Killing-Yano $p$-forms of
Minkowski spacetime. In particular, we will show that all the charges
constructed in Ref.~\cite{Benedetti:2023ipt} are associated to these
mathematical objects. We present our conclusions and directions for future
work in Section~\ref{sec-conclusion}.

\section{Definition of the charges and their relations}
\label{sec-definitions}

In order to illustrate certain ideas about duality in the context of
generalized charges in QFT, the authors of Ref.~\cite{Benedetti:2023ipt}
(based on previous work and ideas in Ref.~\cite{Benedetti:2021lxj} and
generalizing the 4-dimensional case studied in
Refs.~\cite{Benedetti:2022zbb,Hinterbichler:2022agn}) proposed a construction
of all possible 2- and $(d-2)$-form charges of pure $d$-dimensional Einstein
gravity linearized over Minkowski spacetime. The construction uses the
Cartesian coordinates of the Minkowski background, which obscures their
geometrical and physical meaning and their properties.  In order to study them
it is necessary to rewrite them in a fully coordinate- and frame-independent
form first.

It is not difficult to see that all those charges are contractions of the
linearized Riemann tensor (treated as a 2-form) and its dual (treated as a
$(d-2)$-form) with different 2-forms whose properties, combined with those of
the linearized Riemann tensor\footnote{One only needs to use the same Bianchi
  identities that the full non-linear Riemann tensor satisfies.}, guarantee
the closedness of the charges.

In order to rewrite these charges in a fully coordinate- and frame-independent
way, it is simpler to work first in the context of the non-linear theory and
linearize later on.  Thus, it is natural to consider the following 2- and
$(d-2)$-forms

\begin{subequations}
  \begin{align}
    \label{eq:Qsigmadef}
  \mathbf{Q}[\sigma]
  & \equiv
    R^{ab}\sigma_{ab}\,,
    \\
    & \nonumber \\
    \label{eq:Qtildesigmadef}
  \tilde{\mathbf{Q}}[\sigma]
  & \equiv
    \star R^{ab}\sigma_{ab}
    =
    \star \mathbf{Q}[\sigma]\,,
  \end{align}
\end{subequations}

\noindent
where $R^{ab}$ is the Lorentz curvature 2-form defined in
Eqs.~(\ref{eq:curvaturedef}), $\star R_{ab}$ is its Hodge dual $(d-2)$-form
defined in  (\ref{eq:dualcurvature}), and $\sigma_{ab}$ are the components
of a 2-form $\sigma$

\begin{equation}
  \label{eq:sigmadef}
  \sigma
  =
  \tfrac{1}{2}\sigma_{\mu\nu}dx^{\mu}\wedge dx^{\nu}
  =
  \tfrac{1}{2}\sigma_{ab}e^{a}\wedge e^{b}\,.
\end{equation}

Before we consider the conservation of these charges, some comments are in
order:

\begin{enumerate}
  
\item As we are going to see, the conservation of these charges only depends
  on the Bianchi identities and/or on the vacuum Einstein equations. Thus, we
  can replace the Riemann tensor by any other tensor sharing similar
  properties. That is the case of the Riemann tensor linearized over Minkowski
  spacetime, which we will consider in Section~\ref{sec-AD}. 

\item Some of these charges may be total derivatives. At the classical level,
  when, for instance $\mathbf{Q}[\sigma]=d \mathbf{X}[\sigma]$ for some 1-form
  $\mathbf{X}[\sigma]$, integrating $\mathbf{Q}[\sigma]$ over a close
  2-dimensional surface will always give zero. However, one can integrate
  $\mathbf{X}[\sigma]$ over closed curves to get non-vanishing values of
  charges that may be associated to strings, for instance. At the classical
  level, these charges are well defined as long as $\mathbf{X}[\sigma]$ is
  invariant up to total derivatives under the local symmetries of the
  theory. If we are dealing with linearized gravity, this includes the spin-2
  gauge transformations.

  From the point of view of
  Refs.~\cite{Benedetti:2021lxj,Benedetti:2022zbb,Hinterbichler:2022agn,Benedetti:2023ipt}
  and in the context of linearized gravity, though, the relevant objects are
  the local operators $\mathbf{Q}[\sigma]$ and $\mathbf{X}[\sigma]$,
  $\mathbf{Q}[\sigma]$ is strictly gauge-invariant under the spin-2 gauge
  transformations because it depends on the gauge-invariant linearized Riemann
  tensor. $\mathbf{X}[\sigma]$, however, may or may not be a function of that
  tensor. When it is not, it will not be strictly gauge invariant and it
  should not be taken into account.

  Since we are only interested in the classical charges, we will not consider
  these aspects and we will consider the charges obtained by integrating a
  1-form $\mathbf{X}[\sigma]$ invariant up to total derivatives as a
  well-defined charge.

\item As it turns out, in the context of pure Einstein gravity, these charges
  can be considered as the on-shell\footnote{We will denote identities which
    only hold on-shell with $\doteq$.} expressions of the more general charges
  that include terms proportional to the Ricci scalar and the Ricci 1-form,
  considered by Kastor and Traschen in Ref.~\cite{Kastor:2004jk}:

  \begin{subequations}
    \begin{align}
      \label{eq:QKTsigmadef}
  \mathbf{Q}_{KT}[\sigma]
  & \equiv
\imath_{b}\imath_{a}\left[R^{ab}\wedge \sigma\right]
=
  \left(R^{ab}\sigma_{ab}
    -2 \imath_{a}R^{ab}\wedge \imath_{b}\sigma
    + \imath_{b}\imath_{a}R^{ab} \sigma\right)
    \doteq   \mathbf{Q}[\sigma]\,,
    \\
      & \nonumber \\
      \label{eq:QKTtildesigmadef}
  \tilde{\mathbf{Q}}_{KT}[\sigma]
      & \equiv
        \star\mathbf{Q}_{KT}[\sigma]
            \doteq   \tilde{\mathbf{Q}}[\sigma]\,,
    \end{align}
  \end{subequations}

  These charges were not considered in Ref.~\cite{Benedetti:2023ipt} because,
  on-shell, they are equivalent to those in Eqs.~(\ref{eq:Qsigmadef}) and
  (\ref{eq:Qtildesigmadef}) but they can be relevant because it can be shown
  that $\tilde{\mathbf{Q}}_{KT}[\sigma]$ is conserved off-shell when $\sigma$
  is a Killing-Yano 2-form (KY2F) \cite{Kastor:2004jk} while its on-shell
  equivalent $\tilde{\mathbf{Q}}[\sigma]$ is only conserved on-shell
  \cite{Benedetti:2023ipt}. In the context of our general exploration of the
  possible conserved charges it is natural to consider their duals
  $\mathbf{Q}_{KT}[\sigma]$ as well.

\item The above charges Eqs.~(\ref{eq:QKTsigmadef}) and
  (\ref{eq:QKTtildesigmadef}) are particular cases of the $p$- and
  $(d-p)$-form charges defined by \cite{Kastor:2004jk}

  \begin{subequations}
    \begin{align}
      \label{eq:QKTpsigmadef}
  \mathbf{Q}_{KT}[\sigma^{(p)}]
  & \equiv
\imath_{b}\imath_{a}\left[R^{ab}\wedge \sigma^{(p)}\right]\,,
    \\
      & \nonumber \\
      \label{eq:QKTtildepsigmadef}
\tilde{\mathbf{Q}}_{KT}[\sigma^{(p)}]
      & \equiv
        \star\mathbf{Q}_{KT}[\sigma^{(p)}]\,.
    \end{align}
  \end{subequations}

\noindent
where $\sigma^{(p)}$ is a $p$-form ($p>0$)

\begin{equation}
\sigma^{(p)}
=
\frac{1}{p!}
\sigma^{(p)}_{\mu_{1}\cdots \mu_{p}}dx^{\mu_{1}}\wedge \cdots \wedge dx^{\mu_{p}}\,.
\end{equation}

\noindent
The charges $\tilde{\mathbf{Q}}_{KT}[\sigma^{(p)}]$ were shown to be conserved
off-shell for $\sigma^{(p)}$s which are Killing-Yano $p$-forms (KY$p$Fs) in
Ref.~\cite{Kastor:2004jk}.

Notice that for $p=1$

\begin{equation}
  \label{eq:p=1case}
    \mathbf{Q}_{KT}[\sigma^{(1)}]
    =
    -2\sigma^{(1)\, a}G_{ab}e^{b}\,.
\end{equation}
  
In its turn, this leads us to consider the $p$-form generalization of the
charges Eqs.~(\ref{eq:Qsigmadef}) and (\ref{eq:Qtildesigmadef})

  \begin{subequations}
  \begin{align}
    \label{eq:Qpsigmadef}
  \mathbf{Q}[\sigma^{(p)}]
  & \equiv
    R^{ab}\wedge \imath_{b}\imath_{a}\sigma^{(p)}\,,
    \\
    & \nonumber \\
    \label{eq:Qtildepsigmadef}
  \tilde{\mathbf{Q}}[\sigma^{(p)}]
  & \equiv
\star \mathbf{Q}[\sigma^{(p)}]\,,
  \end{align}
\end{subequations}

\noindent
which are equivalent to the former on-shell in pure Einstein
gravity:

\begin{subequations}
      \label{eq:QKTsigmaversusQsigmaonshell}
    \begin{align}
  \mathbf{Q}_{KT}[\sigma^{(p)}]
      & \doteq
          \mathbf{Q}[\sigma^{(p)}]\,,
    \\
      & \nonumber \\
  \tilde{\mathbf{Q}}_{KT}[\sigma^{(p)}]
      & \doteq
  \tilde{\mathbf{Q}}[\sigma^{(p)}]\,.
    \end{align}
\end{subequations}

Observe the definitions Eqs.~(\ref{eq:Qpsigmadef}) and
(\ref{eq:Qtildepsigmadef}) require $p\geq 2$ to be non-trivial while the
Kastor-Traschen charges can be defined for all $p$, but vanish on-shell for
$p<2$. We will not write a superscript for $p=2$.

\end{enumerate}

Thus, it may seem that we will have to study the charges
$\mathbf{Q}[\sigma^{(p)}]$ and $\mathbf{Q}_{KT}[\sigma^{(p)}]$ and their duals
for all $p$, considering later the special case $p=2$. However, the following
lemma will allow us to place limits to the apparent proliferation of charges
and focus only on the simplest of them, namely the $\mathbf{Q}[\sigma^{(p)}]$s
and their duals.

\vspace{.3cm}
\noindent
\textbf{Lemma:} For $p$- and $(d-p)$-form parameters related by

\begin{equation}
  \tilde{\sigma}^{(d-p)}
  =
  \star\sigma^{(p)}\,,  
\end{equation}

\noindent
the charges (\ref{eq:QKTtildepsigmadef}) and the charges
Eqs.~(\ref{eq:Qpsigmadef}) are related by

\begin{equation}
  \label{eq:dualityQKTQ}
  \tilde{\mathbf{Q}}_{KT}[\sigma^{(p)}]
  =
  \mathbf{Q}[\tilde{\sigma}^{(d-p)}]\,.
\end{equation}

\vspace{.3cm}
\noindent
\textbf{Proof:} Using the property
Eq.~(\ref{eq:innerproductsversuswedgeproducts}) and the components of the
curvature 2-form, we get

\begin{equation}
  \begin{aligned}
    \tilde{\mathbf{Q}}_{KT}[\sigma^{(p)}]
    & =
    \star\imath_{b}\imath_{a}\left[R^{ab}\wedge \sigma^{(p)}\right]
    \\
    & \\
    & =
    e_{a}\wedge e_{b}\wedge \star \left[R^{ab}\wedge \sigma^{(p)}\right]
    \\
    & \\
    & =
    \tfrac{1}{2} R_{cd}{}^{ab} e_{a}\wedge e_{b}\wedge \star
    \left[e^{c}\wedge e^{d}\wedge
      \sigma^{(p)}\right]
    \\
    & \\
    & =
    R_{cd}\wedge \star \left[e^{c}\wedge e^{d}\wedge \sigma^{(p)}\right]\,,
  \end{aligned}
\end{equation}

\noindent
where we have used the Bianchi identity Eq.~(\ref{eq:Bianchi5}) in the last
step.  In order to use Eq.~(\ref{eq:innerproductsversuswedgeproducts}) again,
we Hodge-dualize twice $\sigma^{(p)}$ taking into account
Eq.~(\ref{eq:starsquared})

\begin{equation}
  \begin{aligned}
    \tilde{\mathbf{Q}}_{KT}[\sigma^{(p)}] & = (-1)^{p(d-p)}\mathrm{det}(\eta)
    R_{cd}\wedge \star \left[e^{c}\wedge e^{d}\wedge \star^{2}
      \sigma^{(p)}\right]
    \\
    & \\
    & =
    (-1)^{p(d-p)}\mathrm{det}(\eta) R_{cd}\wedge\star^{2}
    \left[\imath_{d}\imath_{c}\star \sigma^{(p)}\right]
    \\
    & \\
    & =
    R_{cd}\wedge \left[\imath_{d}\imath_{c}\star\sigma^{(p)}\right]
    \\
    & \\
    & =
    \mathbf{Q}[\tilde{\sigma}^{(d-p)}]\,,
  \end{aligned}
\end{equation}

\noindent
\textit{quod erat demonstradum}.

This is an important result which, among other things, together with
Eqs.~(\ref{eq:QKTsigmaversusQsigmaonshell}) implies the on-shell relation

\begin{equation}
  \label{eq:onshellduality}
  \tilde{\mathbf{Q}}[\sigma^{(p)}]
  \doteq
  \mathbf{Q}[\tilde{\sigma}^{(d-p)}]\,.
\end{equation}

\noindent
For $p=2$ in $d=4$ it relates all the $\mathbf{Q}[\sigma]$ charges to their
duals on-shell.

We conclude that it is enough to consider the conservation of the
$\mathbf{Q}[\sigma^{(p)}]$s for all values of $p\geq 2$.

\section{The conservation of $\mathbf{Q}[\sigma^{(p)}]$}
\label{sec-conservationofQ}

We are going to study the conservation of the $p$-form charges
$\mathbf{Q}[\sigma^{(p)}]$ defined in Eq.~(\ref{eq:Qpsigmadef}) assuming that
$\sigma^{(p)}$ is a conformal Killing-Yano $p$-form (CKY$p$F).

By definition, a CKY$p$F satisfies an equation of the form (the CKY$p$F
equation)\footnote{The following definitions and properties of CKY$p$Fs can be
  found in the review Ref.~\cite{Frolov:2017kze} and referenced therein.}

\begin{equation}
  \label{eq:CKYpFequation}
  \mathcal{D}_{a}\sigma^{(p)}_{b_{1}\cdots b_{p}}
  =
  \frac{1}{p+1}\aleph^{(p+1)}_{ab_{1}\cdots b_{p}}
  +(-1)^{d(p+1)}\frac{p}{(d-p+1)}\eta_{a[b_{1}}\xi^{(p-1)}_{b_{2}\cdots b_{p}]}\,,
\end{equation}

\noindent
for some $(p+1)$- and $(p-1)$-forms $\aleph,\xi$, which we can identify as

\begin{subequations}
  \begin{align}
    \aleph^{(p+1)}
    & =
      d\sigma^{(p)}\,,
    \\
    & \nonumber \\
    \xi^{(p-1)}
    & =
  \star d\star \sigma^{(p)}\,.
  \end{align}
\end{subequations}

If $\aleph^{(p+1)}=0$, the CKY$p$F is a closed CKY$p$F (CCKY$p$F) and, if
$\xi^{(p-1)}=0$, it is a Killing-Yano $p$-form (KY$p$F). If both
$\aleph^{(p+1)}$ and $\xi^{(p-1)}$ vanish, $\sigma^{(p)}$ is a covariantly
constant CKY$p$F (CCCKY$p$F or, better, C3KY$p$F ). This case is a particular
sub-case of the CCKY$p$F and KY$p$F ones.

From the CKY$p$F equation we get



\begin{equation}
  \label{eq:CKYpFequationab}
  \mathcal{D}\imath_{b}\imath_{a}\sigma^{(p)}
  =
\frac{p-1}{p+1} \imath_{b}\imath_{a}\aleph^{(p+1)}\,. 
\end{equation}

Then, using the Bianchi identity Eq.~(\ref{eq:Bianchi1}) and the above
equation we find 

\begin{equation}
    d\mathbf{Q}[\sigma^{(p)}]
    =
     \frac{p-1}{p+1}R^{ab}\wedge \imath_{b}\imath_{a}\aleph^{(p+1)}\,,
\end{equation}

\noindent
which vanishes trivially off-shell when $\sigma^{(p)}$ is a CCKY$p$F. When it
is not closed, we can proceed as follows: we take the components of the dual
of the above expression written in terms of the components of the dual of
$\aleph^{(p+1)}$

\begin{equation}
  \begin{aligned}
    \left(\star d\mathbf{Q}[\sigma^{(p)}]\right)_{c_{1}\cdots c_{d-p-1}}
    & \sim
2 R^{f_{1}f_{2}}{}_{f_{1}f_{2}}\left(\star \aleph^{(p+1)}\right)_{c_{1}\cdots
  c_{d-p-1}}
\\
& \\
& \hspace{.5cm}
    +4(d-p-1)R^{f_{1}f_{2}}{}_{[c_{1}|f_{1}}\left(\star
      \aleph^{(p+1)}\right)_{f_{2}|c_{2}\cdots c_{d-p-1}]}
\\
& \\
& \hspace{.5cm}
+(d-p-1)(d-p-2)    
    R^{f_{1}f_{2}}{}_{[c_{1}c_{2}}\left(\star \aleph^{(p+1)}\right)_{c_{3}\cdots c_{d-p-1}]f_{1}f_{2}}\,.
  \end{aligned}
\end{equation}

The first two terms vanish on-shell for pure Einstein gravity but the third
only vanishes in $d=p+1$ or $d=p+2$ dimensions.

Thus, we have shown that

\begin{enumerate}
\item $d\mathbf{Q}[\sigma^{(p)}]=0$ for CCKY$p$Fs.
\item $d\mathbf{Q}[\sigma^{(p)}]\doteq 0$ for KY$p$Fs only in
  $d=(p+1)$-dimensional spacetimes whose Ricci scalar vanishes.
\item $d\mathbf{Q}[\sigma^{(p)}]\doteq 0$ for KY$p$Fs only in
  $d=(p+2)$-dimensional Ricci-flat spacetimes.
\end{enumerate}

At this point, the following results become relevant to the discussion
\cite{Frolov:2017kze}:

\begin{enumerate}
\item If $\sigma^{(p)}$ is a CKY$p$F, then
  $\tilde{\sigma}^{(d-p)}=\star \sigma^{(p)}$ is a CKY$(d-p)$F.

\item If $\sigma^{(p)}$ is a CCKY$p$F, then
  $\tilde{\sigma}^{(d-p)}=\star \sigma^{(p)}$ is a KY$(d-p)$F and
  vice-versa. This implies that, if $\sigma^{(p)}$ is a C3KY$p$F, then
  $\tilde{\sigma}^{(d-p)}=\star \sigma^{(p)}$ is a C3KY$p$F.

\item The wedge product of a CCKY$p$F and a  CCKY$q$F
  is a  CCKY$(p+q)$. Observe that a CKY1F is the (metric) dual of a
  conformal Killing vector (CKV).
  
\item The maximal number of CCKY$p$Fs is

  \begin{equation}
    \label{eq:maximalnumberofCCKYpFs}
  \frac{(d+1)!}{p!\cdot(d-p+1)!}\,.
\end{equation}

\end{enumerate}

Then, according to our main result Eq.~(\ref{eq:dualityQKTQ}), there is a
one-to-one relation between the off-shell conserved Kastor-Traschen-type
$p$-forms $\tilde{\mathbf{Q}}_{KT}[\tilde{\sigma}^{(d-p)}]$ constructed with
KY$(d-p)$Fs $\tilde{\sigma}^{(d-p)}$ and the off-shell conserved $p$-forms
$\mathbf{Q}[\sigma^{(p)}]$ constructed with CCKY$p$Fs $\sigma^{(p)}$. They
are, actually, identical.

Furthermore, in $d=p+1,p+2$, one can construct one on-shell conserved
Kastor-Traschen-type $p$-form
$\tilde{\mathbf{Q}}_{KT}[\tilde{\sigma}^{(d-p)}]$ with a CCKY$(d-p)$Fs
$\tilde{\sigma}^{(d-p)}$ for each on-shell conserved $p$-form
$\mathbf{Q}[\sigma^{(p)}]$ constructed with a KY$p$Fs $\sigma^{(p)}$. Again,
they are identical.

\section{Abbott-Deser currents and charges}
\label{sec-AD}

Following Ref.~\cite{Abbott:1981ff,Kastor:2004jk} let us
consider metrics $g$ that asymptote to a given metric $\bar{g}$ so that near
infinity we can express them as perturbations $h$ over the background metric
$\overline{g}$, \textit{i.e.}

\begin{equation}
g_{\mu\nu} = \bar{g}_{\mu\nu}+\chi h_{\mu\nu}\,,  
\end{equation}

\noindent
where $\chi^{2}= 16\pi G_{N}^{(d)}$.

Since we have chosen to work with Vielbein we have to study the linearization
of the Vielbein, spin connection and Lorentz curvature tensor in this
formalism first.\footnote{The Abbott-Deser and Kastor-Traschen charges have
  been written in the Vielbein formalism in
  Refs.~\cite{Cebeci:2006hx,Cebeci:2006mc}.}

\subsection{Linearized gravity in the Vielbein formalism}
\label{sec-LinearizedVielbein}

The Vielbein $e^{a}=e^{a}{}_{\mu}dx^{\mu}$ satisfies the relations

\begin{equation}
  \eta_{ab} e^{a}{}_{\mu}e^{b}{}_{\nu}
  =
  g_{\mu\nu}\,,
  \,\,\,\,\,
  \text{and}
  \,\,\,\,\,  
  e^{a}{}_{\mu}e^{b}{}_{\nu}g^{\mu\nu}
  =
  \eta^{ab}\,,
\end{equation}

\noindent
and the background Vielbein field
$\bar{e}^{\,a}=\bar{e}^{\,a}{}_{\mu}dx^{\mu}$ is assumed to satisfy analogous
relations with respect to the background metric $\bar{g}_{\mu\nu}$ and to the
same tangent space metric $\eta_{ab}$.

Then, we define the perturbation of the Vielbein
$ f^{a} = f^{a}{}_{\mu}dx^{\mu}$ by

\begin{equation}
  e^{a} = \bar{e}^{\,a} + \frac{\chi}{2} f^{a} \,.
\end{equation}

\noindent
By definition,

\begin{equation}
    g_{\mu\nu}
     =
    \bar{g}_{\mu\nu}
    +\chi \eta_{ab}\bar{e}^{\,a}{}_{(\mu}f^{b}{}_{\nu)}
    +\mathcal{O}(\chi^{2})\,,
\end{equation}

\noindent
which requires, for consistency

\begin{equation}
  \eta_{ab}\bar{e}^{\,a}{}_{(\mu}f^{b}{}_{\nu)}
  =
  f_{(\mu\nu)}
  =
  h_{\mu\nu} +\mathcal{O}(\chi)\,.
\end{equation}

\noindent
We use the background Vielbein field to convert tangent space indices into
world indices and vice-versa, the background metric to raise and lower world
indices and the flat Minkowski metric to raise and lower tangent space
indices.

Observe that $f_{\mu\nu}$ has an antisymmetric part

\begin{equation}
  \eta_{ab}\bar{e}^{\,a}{}_{[\mu}f^{b}{}_{\nu]}
  =
  f_{[\mu\nu]}
  \equiv
  b_{\mu\nu}\,,
\end{equation}

\noindent
which does not vanish in general.

Expanding the spin connection as

\begin{equation}
  \omega^{a}{}_{b}
  =
  \bar{\omega}^{a}{}_{b}
  +\chi\omega_{L}{}^{a}{}_{b}+\mathcal{O}(\chi^{2})\,,
\end{equation}

\noindent
the first Cartan structure equation with zero torsion leads to







\begin{equation}
  \label{eq:omegaL}
    \omega_{L\, ab}
     =
    \omega_{L\, cab}\bar{e}^{\, c}
    =
    \tfrac{1}{4} \left\{
      \imath_{\bar{c}}\imath_{\bar{a}}\bar{\mathcal{D}}f_{b}
      -\imath_{\bar{a}}\imath_{\bar{b}}\bar{\mathcal{D}}f_{c}
      +\imath_{\bar{b}}\imath_{\bar{c}}\bar{\mathcal{D}}f_{a} \right\}\bar{e}^{\, c}\,,
\end{equation}

\noindent
where the inner products $\imath_{\bar{a}}$ are taken with the background vector
fields $\bar{e}_{a}=\bar{e}_{a}{}^{\mu}\partial_{\mu}$.  In components, we
have

\begin{equation}
  \bar{\mathcal{D}}f_{b}
  =
  \bar{\mathcal{D}}_{a}f_{bc} \bar{e}^{\, a}\wedge \bar{e}^{\, c} \,,
  \,\,\,\,\,
  \Rightarrow
  \,\,\,\,\,
\imath_{\bar{a}}\imath_{\bar{c}}  \bar{\mathcal{D}}f_{b}
  =
  2\bar{\mathcal{D}}_{[c|}f_{b|a]} \,,
\end{equation}

\noindent
and we find that the linearized connection is given by

\begin{equation}
  \label{eq:omegaLcomponents}
  \begin{aligned}
    \omega_{L\, ab}
    & =
    \tfrac{1}{2}
    \left\{
      \bar{\mathcal{D}}_{[a}f_{b]}
     +\bar{\mathcal{D}}_{[a|}f_{c|b]}\bar{e}^{\, c}
     +\bar{\mathcal{D}}f_{[ab]}
     \right\}\,,
  \end{aligned}
\end{equation}

\noindent
or 

\begin{equation}
  \label{eq:omegaLcomponents2}
  \omega_{L\, ab}
  =
    \tfrac{1}{2}
    \left\{
      \bar{\mathcal{D}}_{a}f_{(bc)}
     -\bar{\mathcal{D}}_{b}f_{(ac)}
     +\bar{\mathcal{D}}_{c}f_{[ab]}
     \right\}\bar{e}^{\, c}\,,  
\end{equation}

\noindent
just to show that it does not depend on the symmetric part of $f_{ab}$ only
\cite{Ortin:2015hya}.

The linearized curvature tensor follows from the Palatini identity

\begin{equation}
\label{eq:RiemannL}
    R_{L\, ab}
    =
    \bar{\mathcal{D}}\omega_{L\, ab}
    =
    \tfrac{1}{2}\bar{\mathcal{D}}
        \left\{
      \bar{\mathcal{D}}_{[a}f_{b]}
     +\bar{\mathcal{D}}_{[a|}f_{c|b]}\bar{e}^{\, c}
     +\bar{\mathcal{D}}f_{[ab]}
     \right\}\,,
\end{equation}

\noindent
and its components are

\begin{equation}
\label{eq:RiemannLcomponents}
    R_{L\,cd\, ab}
\stackrel{[cd]\, [ab]}{=}
    \tfrac{1}{2}\bar{\mathcal{D}}_{c}
        \left\{
      \bar{\mathcal{D}}_{a}f_{bd}
     +\bar{\mathcal{D}}_{a}f_{db}
     +\bar{\mathcal{D}}_{d}f_{ab}
     \right\}\,,
\end{equation}

\noindent
where the notation $\stackrel{[cd]\, [ab]}{=}$ indicates that the pairs of
indices $cd$ and $ab$ are antisymmetrized in right-hand side.

\noindent
Then,

\begin{equation}
  \begin{aligned}
    R_{L\, ab}\wedge \bar{e}^{\, a}
    & =
    \tfrac{1}{2}\bar{R}_{b}{}^{c}\wedge f_{c}\,,
  \end{aligned}
\end{equation}

\noindent
after use of the Bianchi identity Eq.~(\ref{eq:Bianchi2}) for the Riemann
tensor of the background spacetime.

\noindent
Furthermore,

\begin{equation}
  \bar{\mathcal{D}}  R_{L}{}^{ab}
  =
  -2 \bar{R}^{\,[a|}{}_{c}\wedge \omega_{L}{}^{c|b]}\,.
  \end{equation}

\noindent
Hence, for flat background spacetime ($\bar{R}^{ab}=0$), we recover the
Bianchi identities

\begin{subequations}
  \begin{align}
    \label{eq:BianchiL1}
      \bar{\mathcal{D}}  R_{L}{}^{ab}
    & =
      0\,,
    \\
    & \nonumber \\
    \label{eq:BianchiL2}
  R_{L\, ab}\wedge \bar{e}^{\, a}
    & =
      0\,,
  \end{align}
\end{subequations}

\noindent
which we have used to construct the conserved charges in the previous section.
and, therefore, we will restrict ourselves to that case from now onwards.

We also have (using Eqs.~(\ref{eq:Bianchidual1}) and (\ref{eq:Bianchidual2}))

\begin{subequations}
  \begin{align}
\bar{\mathcal{D}} \bar{\star} R_{L}{}^{ab}
& =
(-1)^{d-1}\bar{\mathcal{D}}_{d}R_{L\, ab}{}^{cd}\imath_{\bar{c}}\bar{\omega}\,,
\\
& \nonumber \\
\bar{\star} R_{L\, ab}\wedge e^{a}
    & =
      R_{\mathrm{ic}\,L}{}^{e}{}_{b} \imath_{\bar{e}}\bar{\omega}\,,
  \end{align}
\end{subequations}

\noindent
both of which vanish on-shell for pure (linearized) gravity.

Under diffeomorphisms generated by vector fields
$\xi = \bar{\xi}+\frac{\chi}{2}\epsilon$, and local Lorentz transformations
generated by parameters
$\sigma^{ab}=\bar{\sigma}^{\, ab}+\frac{\chi}{2}s^{ab}$ to lowest order in
$\chi$ we find that

\begin{subequations}
  \begin{align}
    \delta \bar{e}^{\, a}
    & =
      -\pounds_{\bar{\xi}} \bar{e}^{\, a}
      +\bar{\sigma}^{\, a}{}_{b}\bar{e}^{\, b}\,,
    \\
    & \nonumber \\
    \delta f^{a}
    & =
      -\pounds_{\bar{\xi}} f^{a}
      +\bar{\sigma}^{\, a}{}_{b}f^{b}
      -\pounds_{\epsilon} \bar{e}^{\, a}+s^{a}{}_{b} \bar{e}^{\, b}\,.
  \end{align}
\end{subequations}

\noindent
Thus, both $ \bar{e}^{\, a}$ and $f^{a}$ transform as 1-forms defined over the
background spacetime under diffeomorphisms generated by the background vector
fields $\bar{\xi}$ and also as Lorentz vectors with respect to the Lorentz
transformations of the tangent space of the background spacetime, generated by
$\bar{\sigma}^{\, ab}$. On top of this, there are gauge symmetries generated
by the vector fields $\epsilon$ and local Lorentz parameters $s^{ab}$ which
act on the $f^{a}$ as

\begin{equation}
  \begin{aligned}
    \delta f^{a}
    & =
    -\pounds_{\epsilon} \bar{e}^{\, a} +s^{a}{}_{b} \bar{e}^{\, b}
    \\
    & \\
    & =
    -\bar{\mathcal{D}}\epsilon^{a}
    +\left(s^{a}{}_{b}-\imath_{\epsilon}\bar{\omega}^{\, a}{}_{b}\right)
    \bar{e}^{\, b}\,.
  \end{aligned}
\end{equation}

Observe that

\begin{subequations}
  \begin{align}
  \delta f_{(ab)}
  & =
    -\bar{\mathcal{D}}_{(b}\epsilon_{a)}\,,
    \\
    & \nonumber \\
  \delta f_{ab}
  & =
      -\bar{\mathcal{D}}_{[b}\epsilon_{a]}
    +\left(s_{ab}-\imath_{\epsilon}\bar{\omega}_{ab}\right)\,,
  \end{align}
\end{subequations}

\noindent
which means that the symmetric part transforms as a spin-2 field while the
antisymmetric part transforms as a Kalb-Ramond 2-form with an additional
St\"uckelberg transformation with a 2-form parameter $s_{ab}$ which can be
used to remove the antisymmetric part of $f_{ab}$, as expected.\footnote{In
  absence of local Lorentz symmetry this is not possible and theories
  constructed in terms of the Vielbein describe a spin-2 and a spin-1
  field. See Section~4.6.1~of Ref.~\cite{Ortin:2015hya}.} We will just fix the
$s_{ab}$ symmetry setting

\begin{equation}
  s_{ab}
  =
  \imath_{\epsilon}\bar{\omega}_{ab}\,,
\end{equation}

\noindent
in order to simplify the gauge transformations of $f^{a}$:

\begin{equation}
  \label{eq:spin2transformations}
  \delta_{\epsilon}f^{a}
  =
  -\bar{\mathcal{D}}\epsilon^{a}\,.
\end{equation}

The linearized connection transforms as a 1-form under the diffeomorphisms of
the background spacetime generated by the vector fields $\bar{\xi}$ and as a
Lorentz tensor under the local Lorentz transformations of the background
spacetime generated by the parameters $\bar{\sigma}^{\, ab}$. Under the spin-2
gauge transformations Eqs.~(\ref{eq:spin2transformations})

\begin{equation}
  \delta_{\epsilon} \omega_{L\, d\, ab}
  =
  -\tfrac{1}{2}
  \left\{ \bar{\mathcal{D}}_{[a}\bar{\mathcal{D}}_{b]}\epsilon_{d}
    +\bar{\mathcal{D}}_{[a|}\bar{\mathcal{D}}_{d}\epsilon_{|b]}
    +\bar{\mathcal{D}}_{d}\bar{\mathcal{D}}_{[a}\epsilon_{b]}
    \right\}\,.
\end{equation}

In Minkowski spacetime this expression can be simplified

\begin{equation}
  \label{eq:deltaepsilonomegaL}
  \delta_{\epsilon} \omega_{L\, ab}
  =
  -\bar{\mathcal{D}}\bar{\mathcal{D}}_{[a}\epsilon_{b]}\,,
\end{equation}

\noindent
and the transformation of the linearized Riemann tensor is

\begin{equation}
  \delta_{\epsilon}R_{L\,ab}
  =
  \bar{\mathcal{D}}\delta_{\epsilon}\omega_{L\, ab}
  =
  -\bar{\mathcal{D}}\bar{\mathcal{D}}
  \bar{\mathcal{D}}_{[a}\epsilon_{b]}
  =
  0\,.
\end{equation}

\subsection{Asymptotic AD charges}
\label{sec-asymptoticADcharges}

As we have stated before, we can simply replace in the definitions of the
charges the Riemann tensor by the Riemann tensor linearized over Minkowski
spacetime using $\sigma^{(p)}$s which are CKY$p$Fs of the background Minkowski
spacetime (now denoted by $\bar{\sigma}^{(p)}$s) and obtain charges which are
conserved under the conditions we discussed for the charges constructed with
full Riemann tensors. Also, we have seen that it is enough to consider the
$\mathbf{Q}[\sigma^{(p)}]$ charges for $p \geq 2$.

Thus, we have to consider the $p$-forms

\begin{equation}
  \mathbf{Q}_{L}[\bar{\sigma}^{(p)}]
  \equiv
  R_{L}^{ab}\wedge \imath_{\bar{b}}\imath_{\bar{a}}\bar{\sigma}^{(p)}\,,
\end{equation}

\noindent
where the inner products $\imath_{\bar{a}}$ are taken with the background vector
fields $\bar{e}_{a}=\bar{e}_{a}{}^{\mu}\partial_{\mu}$ and we assume that the
background satisfies the linearized Einstein equations so that, in particular,

\begin{equation}
  \imath_{\bar{a}} R_{L}^{ab}
  =
  \imath_{\bar{b}}\imath_{\bar{a}} R_{L}^{ab}
  \doteq
  0\,.
\end{equation}

Observe that these charges are equal to the dual linearized KT-type ones

\begin{equation}
  \tilde{\mathbf{Q}}_{KT\, L}[\tilde{\bar{\sigma}}^{(d-p)}]
  \equiv
  \bar{\star} \imath_{\bar{b}}\imath_{\bar{a}}\left[R_{L}^{ab}\wedge
    \tilde{\bar{\sigma}}^{(d-p)}\right]\,.
\end{equation}

We want to obtain the explicit form of these asymptotic AD-type charges and,
in particular, we want to know when they are total derivatives. This means
that we will have to study the duals as well.

If $\bar{\sigma}^{(p)}$ is a CKY$p$F of the background metric

\begin{equation}
  \begin{aligned}
     \mathbf{Q}_{L}[\bar{\sigma}^{(p)}]
    & =   
\bar{\mathcal{D}}\omega_{L}^{ab}\wedge \imath_{\bar{b}}\imath_{\bar{a}}\bar{\sigma}^{(p)}
    \\
    & \\
    & =   
d\left\{\omega_{L}^{ab}\wedge \imath_{\bar{b}}\imath_{\bar{a}}\bar{\sigma}^{(p)}\right\}
+ \frac{p-1}{p+1}\omega_{L}^{ab}\wedge \imath_{\bar{b}}\imath_{\bar{a}}\aleph^{(p+1)}
  \end{aligned}
\end{equation}

\noindent
where we have used Eq.~(\ref{eq:CKYpFequationab}). The last term will always
vanish for CCKY$p$Fs and, due to the relations that we have established, these
are the charges associated to KY$(d-p)$Fs in Ref.~\cite{Kastor:2004jk}.

Since these charges are exact $p$-forms, when we integrate them over compact
$p$-dimensional surfaces, we will always get zero.  As it is well known, in
these cases one has to define the conserved quantities as integrals over
closed $(p-1)$-dimensional surfaces of the background spacetime of the
$(p-1)$-form whose total derivative we have obtained:

\begin{equation}
  \mathcal{Q}[\bar{\sigma}^{(p)}]
  \sim
  \int_{\Sigma^{(p-1)}}
\omega_{L}^{ab}\wedge \imath_{\bar{b}}\imath_{\bar{a}}\bar{\sigma}^{(p)}\,.
\end{equation}

By construction, these charges are invariant under diffeomorphisms and local
Lorentz transformations of the background spacetime but also under spin-2
gauge transformations Eq.~(\ref{eq:spin2transformations}) because the
integrand is invariant up to a total derivative:

\begin{equation}
  \label{eq:gaugeinvariancecharges}
  \begin{aligned}
    \delta_{\epsilon} \mathcal{Q}[\bar{\sigma}^{(p)}] & \sim
    -\int_{\Sigma^{(p-1)}} \bar{\mathcal{D}}\bar{\mathcal{D}}^{a}\epsilon^{b}
    \wedge \imath_{\bar{b}}\imath_{\bar{a}}\bar{\sigma}^{(p)}
    \\
    & \\
    & =
    -\int_{\Sigma^{(p-1)}}\left\{ d\left[\bar{\mathcal{D}}^{a}\epsilon^{b}
        \wedge \imath_{\bar{b}}\imath_{\bar{a}}\bar{\sigma}^{(p)}\right]
      -\frac{p-1}{p+1}\bar{\mathcal{D}}^{a}\epsilon^{b}
      \imath_{\bar{b}}\imath_{\bar{a}}\aleph^{(p+1)} \right\}
    \\
    & \\
    & =
    -\int_{\Sigma^{(p-1)}}d\left[\bar{\mathcal{D}}^{a}\epsilon^{b}
        \wedge \imath_{\bar{b}}\imath_{\bar{a}}\bar{\sigma}^{(p)}\right]
    \\
    & \\
    & =
    0\,.
  \end{aligned}
\end{equation}

\noindent
where we have used Eq.~(\ref{eq:CKYpFequationab}) again, we have assumed that
$\sigma^{(p)}$ is a CCKY$p$F and we have used Stokes' theorem.

When the integration surface $\Sigma^{(p-1)}$ is not closed, the result of the
integral depends on the boundary conditions satisfied by the gauge parameters
$\epsilon^{b}$. This may not be acceptable from the point of view of the
algebraic approach to generalized symmetries in QFT adopted in
Refs.~\cite{Benedetti:2021lxj,Hinterbichler:2022agn,Benedetti:2022zbb,Benedetti:2023ipt}. As
we have stated in the introduction, our approach is purely classical and we
will not discussed the implications of our results in that respect.

Thus, from the purely classical point of view, the off-shell conserved 2-forms
associated to CCKY$2$Fs are exact and actually lead to 1-form charges.  The
off-shell conserved $(d-2)$-forms are associated to non-CCKY$(d-2)$Fs,
including KY$(d-2)$Fs and, in general, they are not exact.

In order to get 2-form charges one may have to consider CCKY$3$Fs, and one
could also consider CCKY$(d-1)$Fs to get $(d-2)$-form charges. We will study
all these possibilities in detail in the next sections, comparing our results
to those in Ref.~\cite{Benedetti:2023ipt}. We start with the simplest case,
$d=4$.

\subsection{The $d=4$ case}

According to the previous discussion, it is not enough to consider the CKY2Fs
of the Minkowski spacetime background to construct 2-form charges: some of
them will give rise to 1-form charges, because our construction gives total
derivatives and, (perhaps) to compensate the problem, some CFKY3Fs and CKVs
will give 2-forms because our construction, again, gives total derivatives.

We start by finding all the CKY2Fs of $d=4$ Minkowski spacetime, to show that
the solutions cover all the 2-forms used in Ref.~\cite{Benedetti:2023ipt}.

\subsubsection{The CKY2Fs of $d=4$ Minkowski spacetime}

The most general CKY2F $\sigma$ satisfies the equation

\begin{equation}
  \label{eq:CKY2Fequation}
  \mathcal{D}\sigma_{ab}
  =
  \tfrac{1}{3}\left(\aleph_{abc}+2\eta_{c[a}\xi_{b]}\right)e^{c}\,, 
\end{equation}

\noindent
where $\xi = \xi_{a}e^{a}$ is a 1-form and
$\aleph=\tfrac{1}{3!}\aleph_{abc}e^{a}\wedge e^{b}\wedge e^{c}$ is a 3-form
which characterize the closedness and co-closedness of the CKY2F $\sigma$. If
we write these equations in the form

\begin{subequations}
  \begin{align}
    d\sigma
    & =
      \aleph\,,
    \\
    & \nonumber \\
     d\star \sigma
    & =
      \star \xi\,,
  \end{align}
\end{subequations}

\noindent
it is evident that the most general solution to Eq.~(\ref{eq:CKY2Fequation})
is characterized by an exact 3-form $\aleph$, a co-exact 1-form $\xi$ and a
covariantly constant (hence closed and co-closed and, therefore, harmonic)
2-form $a = \tfrac{1}{2}a_{ab}e^{a}\wedge e^{b}$.\footnote{This is reminiscent
  of the Hodge decomposition of differential forms.}  On the other hand, the
interchange between $\sigma$ and $\tilde{\sigma}$ corresponds to the
interchange between $\aleph$ and $\star \xi$ and between $a$ and $\star a$.

Since linear combinations of CKY$2$Fs with constant coefficients give
CKY$2$Fs, we may consider separately those which are covariantly constant
$\sigma=c$, those which are closed, $\aleph=0$, and those which are KY$2$Fs,
$\xi=0$. However, not all the CKY$2$Fs can be constructed as linear
combinations of CKY$2$Fs of these three classes, basically because the
integrability conditions allow for solutions which demand for both
non-vanishing $\aleph$ and non-vanishing $\xi$. This class of solutions must
be closed under Hodge duality just as the class of C3KY$2$Fs does.

We are going to see how all this is realized in the case of the 4-dimensional
Minkowski spacetime considered in Ref.~\cite{Benedetti:2023ipt}.  Thus, we use
Cartesian coordinates $x^{\mu}$ and Vielbein
$e^{a}=\delta^{a}{}_{\mu}dx^{\mu}$. In this basis the spin connection vanishes
and $\mathcal{D}=d$.

We are going to consider, in this order, the covariantly constant, the closed
CKY2Fs, the KY2Fs and the case of $\sigma$s which have $\xi\neq 0$ and
$\aleph\neq 0$.

\begin{enumerate}

  
\item The covariantly constant bivectors $\sigma^{ab}= a^{ab}$ are purely
  constant bivectors with 6 independent components.  $\sigma$ is an exact
  2-form and

\begin{equation}
  \sigma
  =
  d \left(\tfrac{1}{2}a_{\mu\nu}x^{\mu}dx^{\nu}\right)\,.
\end{equation}

In other words: the constant $\sigma^{ab}$s are the \textit{Killing bivectors}
or \textit{momentum maps} \cite{Elgood:2020svt} of the vectors that generate
Lorentz transformations $k_{ab}=k_{ab}{}^{\mu}\partial_{\mu}$ with
$k_{ab}{}^{\mu}=\eta_{ab}{}^{\mu}{}_{\nu}x^{\nu}$:

\begin{equation}
  \partial^{a}\left(-a^{cd}k_{cd}{}^{b}\right)
  =
  \partial^{a}\left(-a^{cd}\eta_{cd}{}^{b}{}_{\nu}x^{\nu}\right)
  =
  -a^{cd}\eta_{cd}{}^{ba}
  =
  a^{ab}\,.
\end{equation}

This class of 2-forms is evidently closed under Hodge duality. 

For each of the 6 independent $a$s we get an off-shell conserved charge
$\mathbf{Q}_{L}[a]$ which is exact and which leads to non-trivial 1-form
charges only. Our main result Eq.~(\ref{eq:dualityQKTQ}) applied to the
charges constructed with the linearized Riemann tensors tells us that
$\mathbf{Q}_{L}[a]=\tilde{\mathbf{Q}}_{KT\, L}[\tilde{a}]$ and that this
charge is also a total derivative.

Furthermore, on-shell
$\mathbf{Q}_{L}[a]=\tilde{\mathbf{Q}}_{KT\, L}[\tilde{a}]\doteq
\tilde{\mathbf{Q}}[\tilde{a}]$ will have the same value on-shell and does not
give an independent charge. (This is just Eq.~(\ref{eq:onshellduality})).

Finally, $\tilde{\mathbf{Q}}[\tilde{a}]=\mathbf{Q}_{KT\, L}[a]$ and this last
charge is not independent, either.


\item The CCKY2Fs ($\aleph=0$) satisfy the equation

\begin{equation}
  d\sigma^{ab}
  =
  \tfrac{2}{3}\delta^{[a}{}_{\mu}\xi^{b]} dx^{\mu}\,.
\end{equation}

The integrability condition is

\begin{equation}
  \delta^{[a}{}_{\mu}\delta^{b]}{}_{\nu}d\xi^{\nu}\wedge dx^{\mu}
  =
  0\,,
\end{equation}

\noindent
which is solved by vectors with constant components $\xi^{a}$. These are the
4 Killing vectors that generate translations and their dual 1-forms are
exact: $\xi_{a}e^{a}= d(\xi_{\mu}x^{\mu})$.

Then, redefining $\xi^{a}\rightarrow 3\xi^{a}$ the solutions are of the form

\begin{equation}
  \label{eq:lesstrivialMinkowski}
  \sigma^{ab}
  =
  2\delta^{[a}{}_{\mu}\xi^{b]} x^{\mu}
  \equiv
  b^{ab}\,,
\end{equation}

\noindent
where the components $\xi^{b}$ are constant, (up to constant bivectors which
we have already taken into account). Actually, we can view the 2-form
$b=\tfrac{1}{2}b_{ab}e^{a}\wedge e^{b}$ as the exterior product of (dual of)
the CKVs that generates dilatations $\eta_{\mu\nu}x^{\mu}dx^{\nu}$ and the
(dual 1-form) of the KVs that generate translations

\begin{equation}
  b
  =
  2\left(\tfrac{1}{2}\eta_{\mu\nu}x^{\mu}dx^{\nu}\right)\wedge
  \left( \xi_{\rho}dx^{\rho} \right)
  =
  d\left(x^{2}\right)\wedge d \left( \xi_{\rho}x^{\rho} \right)\,,
\end{equation}

\noindent
and are obviously exact.

The total number of CCKY2Fs (including the C3KY$2$Fs) is $6+4$, in agreement
with the general result Eq.~(\ref{eq:maximalnumberofCCKYpFs}). However, all
these give off-shell conserved 2-form charges which turn out to be
exact. Therefore, one can only define with them 10 non-trivial 1-form charges.


\item Next, let us consider the KY2Fs ($\xi=0$), which satisfy the equation

\begin{equation}
  d\sigma_{ab}
  =
  \tfrac{1}{3} \aleph_{ab\mu}dx^{\mu}\,,
\end{equation}

The integrability condition of this equation reads

\begin{equation}
  \partial_{[\mu}\aleph_{\nu] ab}dx^{\mu}\wedge dx^{\nu}
  =
  0\,,
\end{equation}

\noindent
and its only non-trivial solution is a tensor with constant components, and
the components of $\sigma$ are

\begin{equation}
  \sigma_{ab}
  =
  \tfrac{1}{3}\aleph_{ab\mu}x^{\mu}
  \equiv
  c_{ab}\,.
\end{equation}

There are 4 constant 3-forms in 4 dimensions, and they are dual to constant
1-forms $b$

\begin{equation}
  \aleph_{abc} \sim \varepsilon_{abcd}\xi^{d}\,.  
\end{equation}

This is the duality between the KY2Fs ($c$) and the CCKY2Fs ($b$).

With the 4 KY2Fs $c$ we can construct 4 on-shell conserved 2-form charges
$\mathbf{Q}[c]$ which are not total derivatives in 4 dimensions. 


\item Finally, let us consider the general equation

\begin{equation}
  d\sigma_{ab}
  =
  \tfrac{1}{3}\left(\aleph_{ab\mu}dx^{\mu}
    +2\eta_{\mu[a}\xi_{b]} dx^{\mu}\right)\,,
\end{equation}

\noindent
whose integrability condition is

\begin{equation}
  \left(\partial_{[\mu}\aleph_{\nu]}{}^{ab}
    +2\eta^{[a}{}_{[\nu}\partial_{\mu]}\xi^{b]}\right)dx^{\mu}\wedge dx^{\nu}
  =
  0\,.
\end{equation}

This equation admits solutions which are not combinations of those belonging
to the previous cases (general CKY2Fs) \cite{Benedetti:2023ipt}:

\begin{equation}
  \aleph_{\nu}{}^{ab}
  =
  \eta_{\nu \rho}x^{[\rho}a^{ab]}\,,
  \hspace{1cm}
  \xi^{b}
  =
  \tfrac{1}{3}a^{b}{}_{\mu}x^{\mu}\,,
\end{equation}

\noindent
with a constant, antisymmetric $a^{ab}$, and the solution of the original
equation is just

\begin{equation}
    \sigma^{ab} =
    \tfrac{1}{2}a^{ab}x^{2} +2a^{[a}{}_{\mu}\delta^{b]}{}_{\nu}x^{\mu}x^{\nu}
    \equiv
    d^{ab}\,,
\end{equation}

\noindent
in agreement with Ref.~\cite{Benedetti:2023ipt}. There is an independent
solution for each independent choice of $d^{ab}$, that is 6 in 4 dimensions,
and for each of them there is an on-shell conserved charge $\mathbf{Q}_{L}[d]$
in 4 dimensions.

\end{enumerate}

Thus, in $d=4$ dimensions, using only CKY2Fs we can only construct 10
independent on-shell conserved 2-form charges and no off-shell 2-forms
whatsoever. However we still have to consider the CCKY3Fs.

\subsubsection{The CCKY3Fs of $d=4$ Minkowski spacetime}

CKY3Fs satisfy the equation

\begin{equation}
  \partial_{a}\sigma^{(3)}_{bcd}
  =
  \tfrac{3}{2}\eta_{a[b}\xi^{(2)}_{cd]}\,,
\end{equation}

\noindent
which admits two classes of solutions:

\begin{enumerate}

\item Constant 3-forms ($\xi^{(2)}=0$), of which there are 4 independent in 4
  dimensions, dual to constant vectors

\begin{equation}
  \sigma^{(3)}_{abc}
  =
  \varepsilon_{abcd}\xi^{d}
  \equiv
  f_{abc}\,.  
\end{equation}

These are, actually, the CCKY3Fs dual to the 4 translational KVs (KY1Fs)
$\xi$.  One can construct with them 4 off-shell conserved 3-form charges
$\mathbf{Q}_{L}[f] \doteq \tilde{\mathbf{Q}}_{L}[b]$ which would be the
3-forms considered in \cite{Kastor:2004jk}. They are exact and give rise to 4
off-shell conserved 2-form charges.

\item 3-forms associated to the dual of the Killing vectors $k$ that generate
  Lorentz transformations with constant parameters $a_{ab}$:

\begin{equation}
  \sigma^{(3)}_{abc}
  =
  -3\tilde{a}_{[ab}\eta_{c]\mu}x^{\mu}
  \equiv
  l_{abc}\,.
\end{equation}

Indeed, if $k$ is such a Killing vector

\begin{equation}
  (\star k)^{a}
  =
  \tilde{\sigma}^{\, (1)\, a}
  =
  a^{a}{}_{\mu}x^{\mu}\,.
\end{equation}
On the other hand, these CCKY3Fs can be seen as the exterior product of the
constant CCKY2Fs $a$ and the CKV that generated dilatations.

There are 6 of these and, again they give off-shell conserved 3-form charges
of the type considered in \cite{Kastor:2004jk}
$\mathbf{Q}_{L}[l]=\tilde{\mathbf{Q}}_{KT\, L}[\xi]$ which are exact and give
rise to 6 off-shell conserved 2-form charges.
  
\end{enumerate}

Thus, we find $4+6$ off-shell additional conserved 2-form charges.

It is worth discussing these charges in some more detail, because, upon
integration at infinity, they are the standard gravitational conserved charges
of asymptotically-flat spacetimes.  As we have shown, the CCKY3Fs are dual to
the 10 KY1Fs of the spacetime, that is, to its Killing vectors, which we can
generically denote by $\bar{k}$. Then, using the duality
Eq.~(\ref{eq:dualityQKTQ}) and Eq.~(\ref{eq:p=1case}), we get

\begin{equation}
 \mathbf{Q}[\tilde{\bar{k}}]
  =
 -2\bar{k}^{\, a}G_{L\, ab} \star \bar{e}^{b}\,. 
\end{equation}

This expression vanishes on-shell and, if we are only interested in strictly
gauge-invariant charges constructed with the linearized Riemann tensor as in
Refs.~\cite{Benedetti:2021lxj,Hinterbichler:2022agn,Benedetti:2022zbb,Benedetti:2023ipt},
they should not be considered.  However, as we have shown,\footnote{As it has
  also been shown, for instance, in \cite{Abbott:1981ff}.} this expression is
the total derivative of a 2-form and the integral of this 3-form does not
necessarily vanish on-shell and is also gauge-invariant
(\ref{eq:gaugeinvariancecharges}).

\begin{equation}
  \begin{aligned}
    \mathcal{Q}[\tilde{\bar{k}}]
    & \sim
    \int_{\Sigma^{(2)}}
    \varepsilon_{abcd}\omega_{L\, e}{}^{ab}\bar{k}^{\,d} \bar{e}^{\,e}\wedge
    \bar{e}^{\,c}
    \\
    & \\
    & =
  -\tfrac{1}{6}  \int_{\Sigma^{(2)}}
  \left\{2\omega_{L\, c}{}^{ca}\bar{k}^{\,b}
  +\omega_{L\, c}{}^{ab}\bar{k}^{\,c}\right\}\imath_{\bar{a}}\imath_{\bar{b}}\bar{\omega}\,,
    \\
    & \\
    & =
  -\tfrac{1}{24}  \int_{\Sigma^{(2)}}
  \left\{2      \left[
      \bar{\mathcal{D}}_{c}f^{ac}
      +\bar{\mathcal{D}}_{c}f^{ca}
     -\bar{\mathcal{D}}^{a}f^{c}{}_{c}
     -\bar{\mathcal{D}}^{a}f_{c}{}^{c}
     +\bar{\mathcal{D}}_{c}f^{ca}
     -\bar{\mathcal{D}}_{c}f^{ac}
     \right]
\bar{k}^{\,b}
  \right.
  \\
  & \\
  & \hspace{.5cm}
  \left.
  +\left[
      \bar{\mathcal{D}}^{a}f^{b}{}_{c}
      +\bar{\mathcal{D}}^{a}f_{c}{}^{b}
     -\bar{\mathcal{D}}^{b}f^{a}{}_{c}
     -\bar{\mathcal{D}}^{b}f_{c}{}^{a}
     +\bar{\mathcal{D}}_{c}f^{ab}
     -\bar{\mathcal{D}}_{c}f^{ba}
     \right]
\bar{k}^{\,c}\right\}\imath_{\bar{a}}\imath_{\bar{b}}\bar{\omega}
    \\
    & \\
    & =
  -\tfrac{1}{24}  \int_{\Sigma^{(2)}}
  \left\{2      \left[
      2\bar{\mathcal{D}}_{c}f^{ca}
     -\bar{\mathcal{D}}^{a}f^{c}{}_{c}
     -\bar{\mathcal{D}}^{a}f_{c}{}^{c}
     \right]
\bar{k}^{\,b}
  \right.
  \\
  & \\
  & \hspace{.5cm}
  \left.
  +\left[
      2\bar{\mathcal{D}}^{a}f^{b}{}_{c}
      +2\bar{\mathcal{D}}^{a}f_{c}{}^{b}
     +2\bar{\mathcal{D}}_{c}f^{ab}
     \right]
\bar{k}^{\,c}\right\}\imath_{\bar{a}}\imath_{\bar{b}}\bar{\omega}
    \\
    & \\
    & =
  -\tfrac{1}{12}  \int_{\Sigma^{(2)}}
  \left\{\left[
      2\bar{\mathcal{D}}_{c}f^{ca}
     -\bar{\mathcal{D}}^{a}f^{c}{}_{c}
     -\bar{\mathcal{D}}^{a}f_{c}{}^{c}
     \right]
\bar{k}^{\,b}
  \right.
  \\
  & \\
  & \hspace{.5cm}
  \left.
  +\left[
      \bar{\mathcal{D}}^{a}f^{b}{}_{c}
      +\bar{\mathcal{D}}^{a}f_{c}{}^{b}
     +\bar{\mathcal{D}}_{c}f^{ab}
     \right]
\bar{k}^{\,c}\right\}\imath_{\bar{a}}\imath_{\bar{b}}\bar{\omega}\,.
  \end{aligned}
\end{equation}

If we use the gauge symmetry to eliminate the antisymmetric part of $f_{ab}$
and identifying $f_{ab}=2h_{ab}$ this expression simplifies further

\begin{equation}
    \mathcal{Q}[\tilde{\bar{k}}]
     \sim
  -\tfrac{1}{3}  \int_{\Sigma^{(2)}}
  \left\{
      \bar{k}^{\,b}\bar{\mathcal{D}}_{c}h^{ca}
     -\bar{k}^{\,b}\bar{\mathcal{D}}^{a}h_{c}{}^{c}
      +\bar{k}^{\,c}\bar{\mathcal{D}}^{a}h^{b}{}_{c}
     \right\}
   \imath_{\bar{a}}\imath_{\bar{b}}\bar{\omega}\,.
\end{equation}

In Cartesian coordinates $x^{\mu}$ and in the Vielbein basis
$\bar{e}^{\, a}=\delta^{a}{}_{\mu}dx^{\mu}$
($\bar{\mathcal{D}}_{a}=\partial_{a}$) and for the timelike Killing vector
$\bar{k}= \partial_{0}$

\begin{equation}
  \begin{aligned}
    \mathcal{Q}[\tilde{\bar{k}}]
    & \sim
    \tfrac{1}{6} \int_{\Sigma^{(2)}}
    \left\{
      \partial^{i}h^{j}{}_{j}
      -\partial^{j}h^{i}{}_{j}
 \right\}\varepsilon_{ikl}dx^{k}\wedge dx^{l}\,,
  \end{aligned}
\end{equation}

\noindent
which is, up to adequate normalization, the ADM mass
\cite{Arnowitt:1962hi}. 

The charges associated to the rest of the KVs of Minkowski spacetime give the
other 9 conserved quantities that characterize asymptotically-flat spacetimes.

The overall situation is summarized in Table~1.

\begin{sidewaystable}
\label{tab:d=4table}
\begin{center}
\begin{tabular}{||c|c|c|c|c|c|c||}
\hline\hline
 & BBM & Here & KT & exact? & on/off-shell? & \# \\
  \hline\hline
  & & & & & & \\
  i & $A$ & $\tilde{\mathbf{Q}}_{L}[a]$ &
$\mathbf{Q}_{KT\, L}[\tilde{a}]
\doteq \mathbf{Q}_{L}[\tilde{a}]$ &
yes & on-shell & 6 \\
  \hline
  & & & & & & \\
  ii & $B$ & $\tilde{\mathbf{Q}}_{L}[b]$ &
$\mathbf{Q}_{KT\, L}[c]
\doteq \mathbf{Q}_{L}[c]$ &
 & on-shell & 4 \\
  \hline
  & & & & & & \\
  iii & $C$ & $\tilde{\mathbf{Q}}_{L}[c]$ &
$\mathbf{Q}_{KT\, L}[b]
\doteq \mathbf{Q}_{L}[b]$ &
yes & on-shell & 4 \\
  \hline
  & & & & & & \\
  iv & $D$ & $\tilde{\mathbf{Q}}_{L}[d]$ &
$\mathbf{Q}_{KT\, L}[\tilde{d}]
\doteq \mathbf{Q}_{L}[\tilde{d}]$ &
 & on-shell & 6 \\
  \hline\hline
  & & & & & & \\
  v & $\star A$ & $\mathbf{Q}_{L}[a]$ &
$\tilde{\mathbf{Q}}_{KT\, L}[\tilde{a}]
\doteq \tilde{\mathbf{Q}}_{L}[\tilde{a}]$ &
yes & off-shell & 6 \\
  \hline
  & & & & & & \\
  vi & $\star B$ & $\mathbf{Q}_{L}[b]$ &
$\tilde{\mathbf{Q}}_{KT\, L}[c]
\doteq \tilde{\mathbf{Q}}_{L}[c]$ &
yes & off-shell & 4 \\
  \hline
  & & & & & & \\
  vii & $\star C$ & $\mathbf{Q}_{L}[c]$ &
$\tilde{\mathbf{Q}}_{KT\, L}[b]
\doteq \tilde{\mathbf{Q}}_{L}[b]$ &
 & on-shell & 4 \\
  \hline
  & & & & & & \\
  viii & $\star D$ & $\mathbf{Q}_{L}[d]$ &
$\tilde{\mathbf{Q}}_{KT\, L}[\tilde{d}]
\doteq \tilde{\mathbf{Q}}_{L}[\tilde{d}]$ &
&  on-shell &6      \\
  \hline\hline
  & & & & & & \\
  ix &  & $\mathbf{Q}_{L}[f]$ &
$\tilde{\mathbf{Q}}_{KT\, L}[\tilde{f}]
\doteq \tilde{\mathbf{Q}}_{L}[\tilde{f}]$ &
yes & off-shell & 4 \\
  \hline
  & & & & & & \\
  x &  & $\mathbf{Q}_{L}[l]$ &
$\tilde{\mathbf{Q}}_{KT\, L}[\tilde{l}]
\doteq \tilde{\mathbf{Q}}_{L}[\tilde{l}]$ &
yes & off-shell & 6 \\
\hline\hline
\end{tabular}
\end{center}
\caption{\small In this table we represent all the charges that can be
  constructed with the CKY2Fs of 4-dimensional Minkowski spacetime using the
  linearized Riemann tensor. In the second column we write the charge as it is
  referred to in Ref.~\cite{Benedetti:2023ipt} (BBM). The form which is
  actually integrated is the dual. In the third column we write the same
  charge in our notation and in the fourth we write the Kastor-Traschen-type
  charge \cite{Kastor:2004jk} which is strictly equivalent to the other two
  upon use of Eq.~(\ref{eq:dualityQKTQ}) and, next to it, the charge which is
  equivalent to it on-shell. In the next columns we indicate whether the
  charge is exact, conserved on- or off-shell and the number of charges of
  that kind. The charges in the rows i and v, iv and viii, ii and vii and iii
  and vi are related by duality and have the same values. Only half of them
  are independent and a half of that half are exact. Thus, only the pairs ii-v
  and iv-vii are independent, not exact, 2-form charges and they all turn out
  to be conserved on-shell. The rows ix and x describe exact, off-shell
  conserved charges associated to 3-forms which give rise to 2-form charges.}
\end{sidewaystable}

\subsection{The  $d=5$ case}
\label{sec-Minkowskid=5}

Instead of considering the arbitrary $d>4$ case, we will just consider the
$d=5$ case which already exhibits the main features of the general case and is
somewhat easier to handle. Using the dualities and on-shell relations we have
uncovered, it is enough to focus on the $\mathbf{Q}_{L}[\sigma^{(p)}]$ charges
with $p=2,3,4$ if we are just interested in 2- and 3-form conserved,
independent and nontrivial charges.

It is easy to see that the CKY2Fs do not give any: for the constant (10) and
closed (5) CKY2Fs the $\mathbf{Q}_{L}[\sigma^{(2)}]$s are exact and only give
1-form charges. The (10) KY2Fs and the (10) CKY2Fs which are not closed do not
give any conserved charges in 5 dimensions.

The (10) constant and (10) closed CKY3Fs give the off-shell conserved
nontrivial 2-form charges studied in Ref.~\cite{Kastor:2004jk} while the (10)
KY3Fs and the (10) CKY3Fs which are not closed do give on-shell conserved
charges in 5 dimensions.

Finally, the  (5) constant and (10) closed  CKY4Fs give additional off-shell
conserved nontrivial 3-form charges while the KY4Fs and the CKY4Fs which are
not closed do not give any conserved charges in 5 dimensions. The existence of
the off-shell conserved 3-form charges does not follow the pattern of on-shell
conserved $(d-2)$-forms and off-shell conserved 2-forms. However, since the
CCKY4Fs are dual to the KVs, these are the conventional gravitational charges
of asymptotically-flat 5-dimensional spacetimes, as we showed in the
4-dimensional case.

\section{Conclusions}
\label{sec-conclusion}

In this paper we have managed to relate and extend the definitions of the
conserved charges of gravitating systems made in
Refs.~\cite{Benedetti:2021lxj,Benedetti:2022zbb,Hinterbichler:2022agn,Benedetti:2023ipt}
and Ref.~\cite{Kastor:2004jk}. In particular, we have shown how the
definitions of conserved charges of particles and fields evolving in
``vacuum'' spacetimes admitting conformal Killing-Yano $p$-forms can be
extended to definitions of gravitational charges of spacetimes that asymptote
to them. In the construction of the Abbott-Deser-type charges, though, we have
considered only asymptotically-flat spacetimes. However, the main result in
Abbott and Deser's seminal paper Ref.~\cite{Abbott:1981ff} was, precisely, the
extension of these ideas to asymptotically-ADS spacetimes. It is natural to
search for a similar extension of the charges studied here and work in this
direction is currently underway.

\section*{Acknowledgments}

P.M.~and T.O.~would like to thank Pablo Bueno for interesting conversations,
for a preview of Ref.~\cite{Benedetti:2023ipt} and for most valuable comments
on a previous version of this paper. T.O.~would like to thank Glenn Barnich
for useful conversations and hospitality during his visit to ULB. This work
has been supported in part by the MCI, AEI, FEDER (UE) grants
PID2021-125700NB-C21 (``Gravity, Supergravity and Superstrings'' (GRASS)),
PID2021-123021NB-I00 and IFT Centro de Excelencia Severo Ochoa
CEX2020-001007-S and by FICYT through the Asturian grant
SV-PA-21-AYUD/2021/52177. The work of CG-F was supported by the MU grant
FPU21/02222.  TO wishes to thank M.M.~Fern\'andez for her permanent support.

\appendix

\section{Some definitions and identities}
\label{app-someidentities}

In this paper we use the conventions of Ref.~\cite{Ortin:2015hya}
throughout and differential-form language. We collect here the main
definitions and identities used throughout the text.

\subsection{The curvature tensor and Bianchi identities}
\label{app-curvature}

We define the Vielbein and spin connection 1-forms

\begin{equation}
e^{a} = e^{a}{}_{\mu}dx^{\mu}\,,
\hspace{1.5cm}
\omega^{ab} = \omega_{\mu}{}^{ab}dx^{\mu} =-\omega^{ba}\,,  
\end{equation}

\noindent
satisfying

\begin{equation}
\mathcal{D}e^{a}
    \equiv
    de^{a} -\omega^{a}{}_{b}\wedge e^{b}
=
    0\,,
\end{equation}

\noindent
where  $\mathcal{D}$ is the exterior Lorentz-covariant derivative.

The Lorentz curvature 2-form

\begin{equation}
  R_{ab}
  \equiv
  \tfrac{1}{2}
  R_{\mu\nu\, ab}dx^{\mu}\wedge dx^{\nu}
  =
  \tfrac{1}{2}
  R_{cd\, ab} e^{c}\wedge e^{d}\,,
\end{equation}

\noindent
can be defined vie the Ricci identity

\begin{equation}
  \label{eq:Ricciidentitydef}
\mathcal{D}\mathcal{D}\xi^{a}
=
-R^{a}{}_{b}\xi^{b}\,,  
\end{equation}

\noindent
for an arbitrary Lorentz vector $\xi^{a}$, and it is given by

\begin{equation}
  \label{eq:curvaturedef}
  R^{ab}
  =
  d\omega^{ab}-\omega^{a}{}_{c}\wedge \omega^{cb}\,.
\end{equation}

Acting on the Vielbein, we get the Bianchi identity

\begin{equation}
\label{eq:Bianchi2}
  R^{a}{}_{b}\wedge e^{b}
  =
-\tfrac{1}{2}  R_{[cd\, b]}{}^{a} e^{c}\wedge e^{d} \wedge e^{b}
 =
 0\,,
 \,\,\,\,
 \Rightarrow
 \,\,\,\,
 R_{[cd\, b]}{}^{a}
 =
 0\,,
\end{equation}

\noindent
which also implies

\begin{equation}
\label{eq:Bianchi5}
  R_{ab\, cd}
  =
  R_{cd\, ab}\,.
\end{equation}

Acting on $\mathcal{D}\xi^{b}$ for an arbitrary Lorentz vector $\xi^{b}$, we
get

\begin{equation}
\mathcal{D}\mathcal{D}\mathcal{D}\xi^{a}
=
-R^{a}{}_{b}\wedge \mathcal{D}\xi^{b}\,,
\end{equation}

\noindent
but acting on both sides of Eq.~(\ref{eq:Ricciidentitydef}) we get

\begin{equation}
\mathcal{D}\mathcal{D}\mathcal{D}\xi^{a}
=
-\mathcal{D}R^{a}{}_{b}\mathcal{D}\xi^{b}
-R^{a}{}_{b}\wedge \mathcal{D}\xi^{b}\,,
\end{equation}

\noindent
which implies the Bianchi identity

\begin{equation}
    \label{eq:Bianchi1}
    \mathcal{D}R^{ab}
    =0\,,
    \,\,\,\,
    \Leftrightarrow
    \mathcal{D}_{[e}R_{cd]}{}^{ab}
    =
    0\,.
\end{equation}

The Ricci tensor is defined by

\begin{equation}
  R_{\mu\nu}
  \equiv
  R_{\mu\rho\nu}{}^{\rho}\,,
\end{equation}

\noindent
and analogously using tangent space indices:

\begin{equation}
  R_{ab}
  \equiv
  R_{acb}{}^{c}\,,
\end{equation}

\noindent
However, in order to avoid confusion with the curvature tensor 2-form, when
using tangent space indices we will write $R_{\mathrm{ic}\, ab}$.

Contracting the indices $d$ and $b$ of Eq.~(\ref{eq:Bianchi1}) the Bianchi
identity takes the form

\begin{equation}
    \label{eq:Bianchi3}
  2\mathcal{D}_{[e}R_{\mathrm{ic}\, c]}{}^{a}
  +\mathcal{D}_{b}R_{ec}{}^{ab}
  =
  0\,,
\end{equation}

\noindent
and contracting now the indices $e$ and $a$ we arrive at the famous contracted
Bianchi identity

\begin{equation}
\label{eq:Bianchi4}
  \mathcal{D}_{a}G^{ab}
  =
  0\,,
  \,\,\,\,\,
  \text{where}
  \,\,\,\,\,
  G^{ab}
  \equiv
  R_{\mathrm{ic}}{}^{ab}-\tfrac{1}{2}g^{ab}R \,,
\end{equation}

\noindent
is the Einstein tensor.

The Ricci identity for an antisymmetric Lorentz tensor is

\begin{equation}
  \mathcal{D}\mathcal{D}\sigma^{ab}
    =
      2\sigma^{[a|}{}_{c}R^{c|b]}
      =
      \delta_{\sigma}R^{ab}\,,  
\end{equation}

\noindent
where $\delta_{\sigma}$ is an infinitesimal local Lorentz transformation
generated by the parameter $\sigma^{ab}$. The same transformation acts on the
spin connection as

\begin{equation}
  \label{eq:deltasigmaomega}
  \delta_{\sigma}\omega^{ab}
  =
  \mathcal{D}\sigma^{ab}\,.
\end{equation}

We also introduce the Levi-Civita affine connection $\Gamma_{\mu\nu}{}^{\rho}$,
whose components are given by the Christoffel symbols, and the total (Lorentz
and general) covariant derivative, denoted by $\nabla$, which satisfies the
first Vierbein postulate

\begin{equation}
  \nabla e^{a}
  =
  \mathcal{D}e^{a} -\Gamma_{\mu\nu}{}^{a}dx^{\mu}\wedge dx^{\nu}
  =
  0\,.
\end{equation}

As a consequence, the Riemann curvature tensor

\begin{equation}
  R_{\mu\nu\rho}{}^{\sigma}(\Gamma)
  \equiv
  2\partial_{[\mu}\Gamma_{\nu]\rho}{}^{\sigma}
  + 2\Gamma_{[\mu|\lambda}{}^{\sigma} \Gamma_{|\nu]\rho}{}^{\lambda}\,,
\end{equation}

\noindent
is related to the Lorentz curvature tensor we have defined before by

\begin{equation}
  R_{\mu\nu\, ab}(\omega)
  =
  R_{\mu\nu\, \rho\sigma}(\Gamma) e_{a}{}^{\rho}e_{b}{}^{\sigma}\,,  
\end{equation}

\noindent
so we can treat both objects as one and the same.

The Einstein equation in vacuum can be written in terms of the curvature
tensor as follows:

\begin{equation}
  3g^{cde}{}_{abf}R_{cd}{}^{ab}
  =
  -2G_{f}{}^{e}
  =
  0\,.
\end{equation}

In $d=3$ dimensions

\begin{equation}
  g^{cde}{}_{abf}
  =
  \varepsilon^{cde}\varepsilon_{abf}\,,
\end{equation}

\noindent
and the above relation between the Einstein and Riemann tensors can be
inverted

\begin{equation}
  \begin{aligned}
    -2G_{f}{}^{e} g_{egh}{}^{fij}
    & =
    3\varepsilon^{cde}\varepsilon_{egh}
    \varepsilon_{abf}\varepsilon^{fij}R_{cd}{}^{ab}
    \\
    & \\
    & =
    12 g^{cd}{}_{gh} g_{ab}{}^{ij}R_{cd}{}^{ab}
    \\
    & \\
    & =
    12 R_{gh}{}^{ij}\,,
  \end{aligned}
\end{equation}

\noindent
which implies that all the 3-dimensional solutions to the vacuum Einstein
equations are locally flat.

\subsection{On-shell identities involving the dual  of the curvature tensor}
\label{app-dualcurvature}

Here we refer to the Hodge dual of the Lorentz curvature 2-form

\begin{equation}
  \label{eq:dualcurvature}
  \star R^{ab}
  \equiv
  \frac{\varepsilon_{\mu_{1}\cdots \mu_{d-2}}{}^{\rho\sigma} R_{\rho\sigma}{}^{ab}}{2\cdot (d-2)!\sqrt{|g|}}
 dx^{\mu_{1}}\wedge \cdots \wedge dx^{\mu_{d-2}}
  =
  \frac{\varepsilon_{c_{1}\cdots c_{d-2}}{}^{cd}R_{cd}{}^{ab}}{2\cdot (d-2)!}
  e^{c_{1}}\wedge \cdots \wedge e^{c_{d-2}} \,.
\end{equation}

\noindent
as the dual curvature tensor $(d-2)$-form.

Acting with the exterior Lorentz derivative on it, we get 

\begin{equation}
  \label{eq:Bianchidual1}
  \begin{aligned}
    \mathcal{D}\star R_{ab}
    & =
   (-1)^{d-1}\mathcal{D}_{d}R_{ab}{}^{cd}\imath_{c}\omega\,,
  \end{aligned}
\end{equation}

\noindent
where $\omega$ is the $d$-dimensional volume form defined in
Eq.~(\ref{eq:omegadef}) and $\imath_{c}$ stands for the inner product with
the vector field $e_{c} = e_{c}{}^{\mu}\partial_{\mu}$ and where we have used
the identity Eq.~(\ref{eq:d-1es}).  The above expression vanishes on-shell for
pure gravity due to the Bianchi identity Eq.~(\ref{eq:Bianchi3}).

Also, using Eq.~(\ref{eq:d-1es})

\begin{equation}
  \label{eq:Bianchidual2}
  \begin{aligned}
    \star R_{ab}\wedge e^{a}
    & =
      R_{\mathrm{ic}}{}^{e}{}_{b} \imath_{e}\omega\,,
  \end{aligned}
\end{equation}

\noindent
which also vanishes on-shell.

\subsection{Identities involving the $d$-dimensional volume form $\omega$}
\label{app-volumeform}

Another set of identities. First, the definition of the volume form:

\begin{equation}
  \label{eq:omegadef}
  \omega
  \equiv
  e^{0}\wedge e^{1} \wedge \cdots \wedge e^{d-1}
  =
  \frac{(-1)^{d-1}\varepsilon_{a_{1}\cdots a_{d}}}{d!}
  e^{a_{1}}\wedge \cdots \wedge e^{a_{d}}\,.
\end{equation}

The $(-1)^{d-1}$ factor is associated to the mostly minus signature that we
are using.

Then, we can prove the following identities:

\begin{subequations}
  \begin{align}
  \label{eq:d-1es}
  e^{c_{1}}\wedge \cdots \wedge e^{c_{d-1}}
  & =
    (-1)^{d-1}\varepsilon^{c_{1}\cdots c_{d-1}b} \imath_{b}\omega\,,
    \\
    & \nonumber \\
  e^{c_{1}}\wedge \cdots \wedge e^{c_{d-2}}
  & =
  -\tfrac{1}{2}\varepsilon^{c_{1}\cdots c_{d-2}b_{1}b_{2}}
    \imath_{b_{1}}\imath_{b_{2}}\omega\,,
    \\
    & \nonumber \\
    e^{c_{1}}\wedge \cdots \wedge e^{c_{d-3}}
    & =
    -\frac{(-1)^{d-1}}{3!}\varepsilon^{c_{1}\cdots c_{d-3}b_{1}b_{2}b_{3}}
    \imath_{b_{1}}\imath_{b_{2}}\imath_{b_{3}}\omega\,,
    \\
    & \nonumber \\
  \label{eq:volumeformidentity}
    e^{c_{1}}\wedge \cdots \wedge e^{c_{d-n}}
   & =
    \frac{(-1)^{[n/2]}(-1)^{n(d-1)}}{n!}\varepsilon^{c_{1}\cdots c_{d-n}b_{1}\cdots b_{n}}
    \imath_{b_{1}}\cdots \imath_{b_{n}}\omega\,.    
  \end{align}
\end{subequations}

\subsection{Other identities}
\label{app-otheridentities}

With our conventions, for any $p$-form $F^{(p)}$,

\begin{equation}
  \label{eq:starsquared}
\star^{2}F^{(p)} = (-1)^{p(d-p)}\mathrm{det}(\eta)F^{(p)}\,,  
\end{equation}

\noindent
where $\mathrm{det}(\eta)$ is the determinant if the tangent-space metric
$\eta_{ab}$ (which equals $(-1)^{d-1}$ for a $d$-dimensional Lorentzian metric
with mostly minus signature) and, if $p\geq n$,

\begin{equation}
  \label{eq:innerproductsversuswedgeproducts}
\star \imath_{a_{1}}\cdots \imath_{a_{n}}  F^{(p)}
  =
  e^{a_{n}}\wedge \cdots \wedge e^{a_{1}} \wedge \star F^{(p)}\,.
\end{equation}


\end{document}